
%
%

\def\NI{\noindent}
\long\def\UN#1{$\underline{{\vphantom{\hbox{#1}}}\smash{\hbox{#1}}}$}
\magnification=\magstep 1
\overfullrule=0pt
\hfuzz=16pt
\voffset=0.0 true in
\vsize=8.8 true in
   \def\NP{\vfil\eject}
   \baselineskip 20pt
   \parskip 6pt
   \hoffset=0.1 true in
   \hsize=6.3 true in
\nopagenumbers
\pageno=1
\footline={\hfil -- {\folio} -- \hfil}
\headline={\ifnum\pageno=1 \hfill November 1992 \fi}

\hphantom{AA}

\hphantom{AA}

\centerline{\UN{\bf Second-Order Dynamics in the Collective}}

\centerline{\UN{\bf Evolution of Coupled Maps and Automata}}

\vskip 0.4in

\centerline{\bf Philippe--Michel~Binder}
\centerline{\sl Department of Physics, Theoretical Physics}
\centerline{\sl University of Oxford, 1 Keble Road, Oxford OX1 3NP,
United Kingdom}

\vskip 0.4in

\centerline{\bf Vladimir Privman}
\centerline{\sl Department of Physics, Clarkson University,
Potsdam, New York 13699--5820, USA}

\vskip 0.8in

\vfill\vfill

\centerline{\bf ABSTRACT}

We review recent numerical studies and the
phenomenology of spatially synchronized collective states in
many-body dynamical systems. These states exhibit
thermodynamic noise superimposed on the collective, quasiperiodic
order parameter evolution with typically one basic irrational
frequency. We concentrate on the description of the global temporal
properties in terms of second-order difference
equations.

\vfill\vfill

\vskip 0.3in

\NI {\bf PACS:} 05.45.+b, 47.25.Mr, 87.10.+e

\NP

Complex dynamical systems exhibit various types of temporal and
spatial evolution ranging from uniform to chaotic. The mechanisms
leading to self-organization manifested in constant (fixed-point),
limit cycle, persistent-structure, quasiperiodic and other
``collective'' temporal behaviors have been a subject of extensive
investigations.$^{1-11}$ Unfortunately, only for the
simplest families of one-dimensional cellular automata,
significant progress
in the understanding of individual rules has been
achieved.$^{3-7,9}$ A more phenomenological approach is based on
classifying possible collective behaviors based on the experience
with few-parameter systems such as iterative maps in one or more
variables. The usual mean-field type idea is to lump the evolution
of the complex system in one or more ``collective'' order
parameters, for instance the total magnetization of an Ising-spin
automaton, or generally large-wavelength Fourier components of the
spin configuration. This approach, as well as the
phenomenological characterization of the spatio-temporal patterns
observed, are at the heart of most classifications of cellular
automata.$^{1-10}$

Recently, arguments have been advanced$^{8}$ suggesting that
spatially uniform collective evolutions in synchronous-dynamics
complex systems are generically limited to ``trivial'' temporally
constant or period-two cyclic states (at large times). The idea has
been that for higher-period temporal evolution, with integer
periods $3,4,\ldots$ in units of the discrete updating time steps
of the automaton, the domain walls between regions ``out of
phase'' will have generically nonzero velocity. Thus, randomly
generated small droplet regions ``out of phase'' with the
surrounding configuration, are expected to spread irreversibly,
destroying the spatial coherence. Similar arguments were advanced
for quasiperiodic evolution, with domain walls replaced by regions
of phase gradients.$^{11}$

Surprisingly, numerical evidence has been reported indicating the
existence of various ``nontrivial'' self-organized states in
higher-dimensional discrete-spin cellular automata. Originally,
these collective states were observed in certain $d=4,5$
totalistic-rule automata.$^{10}$ This discovery was followed by
extensive numerical studies$^{11-17}$ which both enriched the
family of automata models showing these new collective states, and
also classified various phenomenological properties to be listed
shortly. Moreover, similar collective states were found
recently$^{17-18}$ in continuous-variable coupled-map lattices in
$d=2$ through 6. Collective states were also observed in $1d$
coupled-map lattices;$^{19}$ however, their spatial synchronization
pattern and other properties differ from collective states in $d>1$.

The models studied in Refs.~10--19
showed a great variety of self-organized collective states (see
especially Ref.~17). The most intriguing type is characterized
by the formation of a stable loop-like structure
in the return map, i.e., in the plot of the order parameter
$m(t+1)$ vs.~$m(t)$. Figure~1 illustrates such a plot for the
$d=5$ totalistic automaton rule$^{10,17}$ denoted $R^{5}_{59}$,
in which a site is updated to the value $\sigma_i (t+1)=1$ if the
neighborhood sum over the von~Neumann zone (the site itself and its
$2d$ nearest neighbors) is between 5 and 9 at time $t$, and updated
to $\sigma_i (t+1)=0$ otherwise. The order parameter is defined as

$$ m(t) = \sum_j \sigma_j (t) \bigg / \sum_j 1 \; , \eqno(1) $$

\NI where the sum is over all the lattice points $j$, at which the
spins $\sigma_j (t) = 1 $ or $0$ are located. In Figure~1,
about 1000 data points are shown
after a transient of 5000 time steps has been discarded,
for a system of $12^5$ sites.
The initial condition was a randomly half-filled (with
values 1) lattice; and boundary conditions were periodic. The
large-time average value, $M$, was estimated and subtracted out.
This quantity can be formally defined as

$$ M = \lim_{\tau \to \infty} \left[\tau^{-1}
\sum\limits_{t=t_0}^{t_0+\tau} m (t) \right] \; . \eqno(2) $$

Bifurcation diagrams were studied as well, for coupled-map
lattices for which there is a control parameter.$^{17-19}$
Restricting our discussion to the $d>1$ models, we list several
phenomenological observations based largely on numerical studies.
Firstly, although noisy integral-period states (for example,
period three) were found, noisy quasiperiodic behavior attracted
more interest. Here we focus on such quasiperiodic states,
typically with quasiperiods close to integral values $3,4,\ldots$,
measured, for example, by the winding numbers$^{10-11,17}$ about a
point near the origin in plots like Figure~1.

The noise exemplified by the plot in Figure~1, was found to
decrease ``thermodynamically'' with increasing number of
spins,$^{11}$ according to

$$ {\rm noise\ amplitude} \sim 1 \bigg / \sqrt{ \sum_j 1 } \;\;\; .
\eqno(3) $$

\NI This important observation indicates that the spatial
synchronization mechanism is statistical, short-range in nature,
rather than global, i.e., that the synchronization is
achieved as a local effect, via correlations of adjacent system
parts. Unfortunately, the nature of the spatial self-organization
has not been understood theoretically thus far.

Regarding the temporal evolution, numerical evidence based on
maximum-entropy analyses,$^{15}$ as well as some
phenomenological considerations,$^{16}$ suggest that the apparently
quasiperiodic behavior observed in the new collective states both
in $d=1$ and $d>1$, has in fact periodic time dependence with a
single basic frequency. The quasiperiodicity is thus solely due to
the incommensurability of the associated period with the clock
period $\Delta t = 1$ of the automaton time steps. This
observation can be quantified by the relation

$$ m(t \gg 1 )=M + \sum_{n=1} a_n
\sin \left( {2 \pi n t \over T} + \phi_n \right)
+ {\rm noisy\ contributions} \; . \eqno(4) $$

\NI The noisy contributions have no characteristic frequencies
associated with them, and they are presumably important only for
$d>1$ models, as mentioned earlier.

All the above properties were obtained for fully deterministic
evolution. However, stability of the collective states was
checked$^{11,17}$ by adding a small stochastic element
in the evolution rules
and for a range of initial conditions. Empirically, these states
disappear when the stochastic element increases, or when the initial
conditions significantly depart from the random uniform
configuration. In the latter case the system evolves into a
trivial, typically dilute, disordered-type state. Configurations
with imposed domain walls, and the way the system ``heals'' such
perturbations, were studied as well.$^{11}$

The theoretical situation remains largely unsatisfactory. Indeed,
with the accumulated numerical information, it would be of
interest to understand both the spatial and temporal
self-organization mechanisms of these new collective states. One
step in this direction, focusing on phenomenological conclusions
one can draw from the numerical observation (4), was the proposed
\UN{\sl second-order in time\/} nature of the discrete-time dynamics
of the collective behavior at hand.$^{16}$ In the rest of this
Short Review we summarize the approach of Ref.~16.

Empirically, one observes that the leading oscillatory term in (4)
is usually the dominant contribution:

$$ |a_1| \gg |a_{n>1}| \; . \eqno(5) $$

\NI Such time evolution for $t \gg 1$ can be described by the
linear second-order dynamical rule

$$ \mu (t) = 2\cos (c) \mu(t-1) - \mu(t-2) \; , \eqno(6) $$

\NI where

$$ \mu(t) \equiv m(t) - M \; . \eqno(7) $$

\NI Indeed, the solution of (6) is

$$ \mu (t) = a_1 \sin \left( {2 \pi t \over T } + \phi_1 \right)
\; , \qquad {\rm with} \qquad T={2\pi \over c} \; . \eqno(8) $$

\NI The linear evolution rule (6) only contains one parameter, $c$,
determining the basic frequency. Linear fits of the plots of the
discrete second-order derivative, $\mu(t+1)-2\mu(t)+\mu(t-1)$,
vs.~$\mu (t)$, for several different models,$^{16}$ yielded period
values consistent with both the maximum-entropy and winding-number
estimates. Figure~2 illustrates such a plot for the data used
in Figure~1.

The noisy curve in Figure~2 is not precisely a straight-line, nor
even single-valued. However, it is well known that such
``Lissajous-like'' distortions can be incorporated by allowing for
nonlinear terms in the dynamical rule, which we now rewrite as
follows,

$$ \mu(t+1)-2\mu(t)+\mu(t-1) = -4\sin^2 (c/2) \mu(t) +
\sum\limits_{n=2}^\infty b_n \mu^n(t) + {\rm noisy\ contributions}
. \eqno(9) $$

\NI The form of the contributions leading to noise in finite-size
samples is not clear at present.

In some cases, the single-harmonic ``linear'' second-order
difference equation (6) works quite well. As an illustration,
we present in Figure~3 a plot analogous to Figure~2, but for a
$d=3$ automaton described in Ref.~13: the spin is updated to 1
if the sum over the $3d$ von~Neumann zone is 0 or 5, and set to 0
otherwise. However, examples are known of large
nonlinearity$^{16}$ as well as of ``Lissajous-shaped'' features
obscured by the noise, especially in high dimensions where the
system sizes accessible to numerical simulations are small.

A direct maximum-entropy fit of the parameters of the first six
terms in (4), i.e., including the terms up to $a_{6}$ but
excluding the noisy contribution, yields the $\mu(t)$ sequence
which generates the continuous curve in Figure~4. If we only keep
the leading harmonic, (8), then the elliptical-shaped curve is
obtained; see Figure~4. The general equation of such an
elliptical-shaped single-harmonic return map is

$$ \left[ {\mu (t) \over A} \right] ^2 + \left[ {\mu (t+1) \over A}
\right] ^2 - 2 \cos (c) \left[ {\mu (t) \over A} \right] \left[
{\mu (t+1) \over A} \right] = \sin^2 (c) \; . \eqno(10) $$

While these observations shed little light on the nature of the
spatial self-organization, they indicate that when the noise can
be neglected, i.e., presumably in the infinite-system limit, the
collective states essentially follow a ``mechanistic'' time
evolution which in terms of the average order parameter $m(t)$ is
fully deterministic and \UN{\sl reversible}. The latter property
follows from the second-order (in time) nature of the dynamics. As a
result, the droplet-type arguments which are largely irreversible
in nature, should be incorrect although the mechanism of their
breakdown can be fully illuminated only after the spatial
self-organization is described theoretically. Loosely speaking,
reversibility implies that for any locally disordering mechanism
there should exist a reversed locally ordering mechanism. It seems
likely that neither of these processes involves droplets.

In summary, we surveyed a new emerging field of study of
collective, self-organized behavior in complex systems. The
phenomenological information collected by extensive numerical
simulations suggests that these systems offer an interesting
challenge of theoretical description of spatially
self-organized and synchronized evolution, for which the
conventional equilibrium ``droplet'' stability concepts do not
apply. While the spatial mechanisms remain to be explained, the
time-dependence at least of the global order parameter, could be
quantified phenomenologically by the second-order discrete-time
dynamics outlined in this Short Review.

The authors gratefully
acknowledge discussions and collaboration with B.~Buck and
V.A.~Macaulay. This research was supported in part by the Science
and Engineering Research Council (UK).

\NP

\centerline{\bf REFERENCES}

{\frenchspacing

\item{1.} S.~Wolfram,
Rev.~Mod.~Phys.~{\bf 55}, 601 (1983).

\item{2.} C.H.~Bennett and G.~Grinstein, Phys.~Rev.~Lett.~{\bf 55},
657 (1985).

\item{3.} E.~Jen, J.~Stat.~Phys.~{\bf 43}, 219, 243 (1986).

\item{4.} K.~Sutner, Physica~D~{\bf 45}, 386 (1990).

\item{5.} W.~Li and N.H.~Packard, Complex Systems~{\bf 4}, 281
(1990).

\item{6.} C.G.~Langton, Physica~D{\bf 42}, 12 (1990).

\item{7.} H.A.~Gutowitz, Physica~D{\bf 45}, 136 (1990).

\item{8.} C.H.~Bennett, G.~Grinstein, Y.~He, C.~Jayaprakash and
D.~Mukamel, Phys.~Rev.~A{\bf 41}, 1932 (1990).

\item{9.} P.--M.~Binder, J.~Phys.~A{\bf 24}, L31, 1677 (1991).

\item{10.} H.~Chat\'e and P.~Manneville, Europhys.~Lett.~{\bf 14},
409 (1991).

\item{11.} J.A.C.~Gallas, P.~Grassberger, H.J.~Herrmann and
P.~Ueberholz, Physica~A{\bf 180}, 19 (1992).

\item{12.} B.~Barral, H.~Chat\'e and P.~Manneville,
Phys.~Lett.~A{\bf 163}, 279 (1992).

\item{13.} J.~Hemmingsson, Physica~A{\bf 183}, 255 (1992).

\item{14.} J.A.C.~Gallas, R.~Bourbonnais and H.J.~Herrmann,
Int.~J.~Mod.~Phys.~C{\bf 3}, 347 (1992).

\item{15.} P.--M.~Binder, B.~Buck and V.A.~Macaulay,
J.~Stat.~Phys.~{\bf 68}, 1127 (1992), and unpublished results.

\item{16.} P.--M.~Binder and V.~Privman, Phys.~Rev.~Lett.~{\bf
68}, 3830 (1992).

\item{17.} Review:\ H.~Chat\'e and P.~Manneville,
Prog.~Theor.~Phys.~{\bf 87}, 1 (1992).

\item{18.} H.~Chat\'e and P.~Manneville, Europhys.~Lett.~{\bf 17},
291 (1992).

\item{19.} P.--M.~Binder and V.~Privman, J.~Phys.~A{\bf
25}, L755 (1992).

} \NP

\centerline{\bf FIGURE CAPTIONS}

\NI\hang {\bf Fig.~1:\/} Return map, $m(t+1)-M$ vs.~$m(t)-M$,
for the cellular automaton rule $R^{5}_{59}$,
described in the text: about 1000 time steps are shown. A
transient of the first 5000 time steps was discarded. System size
was $12^5$.

\NI\hang {\bf Fig.~2:\/} Second-order discrete time derivative, the
left-hand side of (9), plotted vs.~$\mu(t)$, for the data shown
in  Figure~1.

\NI\hang {\bf Fig.~3:\/} Second-order time derivative plot similar
to Figure~2, for the 3-dimensional Hemmingsson model,$^{13}$
described in the text.

\NI\hang {\bf Fig.~4:\/} The data of Figure~1 are shown together
with their ``noiseless skeleton'' reconstruction obtained by keeping
the first six oscillatory terms in (4), i.e., up to $a_6$. The
ellipse corresponds to keeping only the leading oscillatory term,
corresponding to relations (6), (8), (10).

\bye